\documentclass[twocolumn,english,pre,aps,showpacs]{revtex4}
\usepackage[T1]{fontenc}
\usepackage[latin1]{inputenc}
\usepackage{graphicx}
\usepackage{amssymb}

\makeatletter

\def\vec#1{\mbox{\boldmath $\mathrm{#1}$}}
\def\Pm{\textrm{Pm}}
\def\Rm{\textrm{Rm}}
\def\RE{\textrm{Re}}
\def\Ha{\textrm{Ha}}

\usepackage{babel}
\makeatother

\begin{document}

\title{Absolute versus convective helical magnetorotational instability
in a Taylor-Couette flow}

\author{J\={a}nis Priede }

\email{J.Priede@coventry.ac.uk}

\affiliation{Applied Mathematics Research Centre, Coventry University, Coventry,
CV1 5FB, United Kingdom}

\author{Gunter Gerbeth}

\affiliation{Department of MHD, Forschungszentrum Dresden-Rossendorf, P.O. Box
510119, D--01314 Dresden, Germany}

\begin{abstract}
We analyze numerically the magnetorotational instability of a Taylor-Couette
flow in a helical magnetic field {[}helical magnetorotational instability
(HMRI)] using the inductionless approximation defined by a zero magnetic
Prandtl number ($\Pm=0).$ The Chebyshev collocation method is used
to calculate the eigenvalue spectrum for small amplitude perturbations.
First, we carry out a detailed conventional linear stability analysis
with respect to perturbations in the form of Fourier modes that corresponds
to the convective instability which is not in general self-sustained.
The helical magnetic field is found to extend the instability to a
relatively narrow range beyond its purely hydrodynamic limit defined
by the Rayleigh line. There is not only a lower critical threshold
at which HMRI appears but also an upper one at which it disappears
again. The latter distinguishes the HMRI from a magnetically modified
Taylor vortex flow. Second, we find an absolute instability threshold
as well. In the hydrodynamically unstable regime before the Rayleigh
line, the threshold of absolute instability is just slightly above
the convective one although the critical wavelength of the former
is noticeably shorter than that of the latter. Beyond the Rayleigh
line the lower threshold of absolute instability rises significantly
above the corresponding convective one while the upper one descends
significantly below its convective counterpart. As a result, the extension
of the absolute HMRI beyond the Rayleigh line is considerably shorter
than that of the convective instability. The absolute HMRI is supposed
to be self-sustained and, thus, experimentally observable without
any external excitation in a system of sufficiently large axial extension. 
\end{abstract}

\pacs{47.20.Qr, 47.65.-d, 95.30.Lz}

\maketitle

\section{Introduction}

The magnetorotational instability (MRI) is known to be able to destabilize
hydrodynamically stable flows by means of an externally imposed magnetic
field as originally shown by Velikhov \cite{Velikhov-1959} and analyzed
in more detail by Chandrasekhar \cite{Chandrasekhar-1960} for cylindrical
Taylor-Couette flow of a perfectly conducting fluid subject to an
axial magnetic field. Three decades later Balbus and Hawley \cite{Balbus-Hawley-1991}
suggested that, in a similar way, the hydrodynamically stable Keplerian
velocity distribution in accretion disks could be rendered turbulent
by the MRI accounting for the formation of stars and entire galaxies
proceeding much faster than it could be accomplished by the viscous
angular-momentum transport alone. Meanwhile this proposition has triggered
not only numerous theoretical and numerical studies \cite{Balbus-Hawley-1998}
but also some experimental efforts as well \cite{Sisan-etal,Nature-2006}.
However, one of the main technical challenges to laboratory MRI is
the magnetic Reynolds number $\Rm$ which is required to be $\sim10$
at least. For a liquid metal with the magnetic Prandtl number $\Pm\sim10^{-5}-10^{-6}$
this translates into a hydrodynamic Reynolds number $\RE=\Rm/\Pm\sim10^{6}-10^{7}$
\cite{Goodman-Ji-2002}. Thus, the base flow on which the MRI is
supposed to be observable may easily be turbulent at such Reynolds
numbers independently of MRI as in the experiment of Sisan \emph{et
al.} \cite{Sisan-etal}. A way to circumvent this problem was proposed
by Hollerbach and R\"udiger \cite{Hollerbach-Ruediger-2005} who
suggested that MRI can take place in the Taylor-Couette flow at $\RE\sim10^{3}$
when the imposed magnetic field is helical rather than purely axial
as in the classical case. The theoretical prediction of this new type
of helical MRI (HMRI) was soon succeeded by a confirming experimental
evidence provided by the so-called PROMISE facility \cite{Rued-apjl,Stefani-etal,Stefani-NJP}.
Nevertheless, these experimental observations have subsequently been
questioned by Liu \emph{et al.} \cite{Liu-etal2006} who find no
such instability in their inviscid theoretical analysis of finite
length cylinders with insulating end caps. They suspect the observed
phenomenon to be a transient growth rather than a self-sustained instability
\cite{Liu-etal2007,Liu2008}.

Indeed, such an interpretation of the HMRI is possible when the analysis
is based only on the conventional linear stability analysis for separate
Fourier modes as done by Hollerbach and R\"udiger \cite{Hollerbach-Ruediger-2005}
following the classical MRI approach. However, there is a principal
difference between the classical and the helical MRIs, namely, the
former is stationary whereas the latter is traveling. It is important
to emphasize that the conventional stability analysis for traveling
waves yields the so-called convective instability threshold at which
the system becomes able to amplify certain externally excited perturbations.
At this threshold the perturbation grows in time only in the frame
of reference moving with its group velocity while it asymptotically
decays in any other frame of reference including the laboratory one
\cite{Landau-87}. Eventually, such a growing while traveling perturbation
reaches the end wall where it is absorbed unless the system is able
to reflect it back. The latter supposes reflection symmetry in the
system which, however, is not the case provided that the magnetic
field is helical. Thus, it is indeed unclear whether the HMRI can
be self-sustained in an ideal Taylor-Couette flow of large but finite
axial extension.

This question is addressed in the second part of the present study
where the absolute HMRI is found to exist besides the convective one
which, in turn, is analyzed in detail in the first part. Note that
the existence of absolute instability is nontrivial as known, for
instance, for the Ponomarenko dynamo \cite{Pono} which has a convective
but no absolute instability threshold \cite{Agris1}. The latter
requires an additional return flow to be included in the original
Ponomarenko model \cite{Agris2}. The distinction between convective
and absolute instabilities is relevant mainly for open flows and unbounded
geometries \cite{Huerre-Monkewitz-1990}. In finite geometries, it
is important to distinguish transiently growing and noise-sustained
perturbations from the self-sustained linear instabilities, which
are always global with the threshold asymptotically approaching from
above that of the absolute instability as the system size increases
\cite{Tobias-etal-1998,Proctor-etal-2000}.

We consider both the convective and the absolute HMRIs in the inductionless
approximation corresponding to $\Pm=0$ that was suggested in our
previous work \cite{Priede-etal-2007}. This approximation, which
leads to a significant simplification of the problem, allows us to
focus exclusively on the HMRI because it does not capture the conventional
MRI \cite{Herron-Goodman-2006}. We show that the HMRI is effective
only in a relatively narrow range of the ratio of rotation rates of
the inner and outer cylinders beyond the limit of purely hydrodynamic
instability defined by the so-called Rayleigh line. For the convective
HMRI, the range of instability is considerably larger for perfectly
conducting cylinders than that for insulating ones. In addition we
find that the HMRI is effective only in a limited range of Reynolds
numbers. Namely, for any unstable mode, there is not only a lower
critical Reynolds number by exceeding which the HMRI sets in but also
an upper one by exceeding which it disappears again. It is this upper
threshold that distinguishes HMRI from a magnetically modified Taylor
vortex flow. Absolute HMRI exists in a significantly narrower range
of parameters than the convective one. In contrast to the convective
HMRI, the absolute one is much less dependent on the conductivity
of the boundaries.

The paper is organized as follows. In Sec. \ref{sec:prob-form} we
formulate the problem using the inductionless approximation. Numerical
results concerning the convective and absolute instability thresholds
for both insulating and perfectly conducting cylinders are presented
in Secs. \ref{sub:Conv} and \ref{sub:Abs-inst}, respectively. Section
\ref{sec:summ} concludes the paper with a summary and a comparison
with experimental results of Stefani \textit{et al.} \cite{Rued-apjl,Stefani-etal,Stefani-NJP}.

\begin{figure}
\begin{centering}
\includegraphics[width=0.3\textwidth]{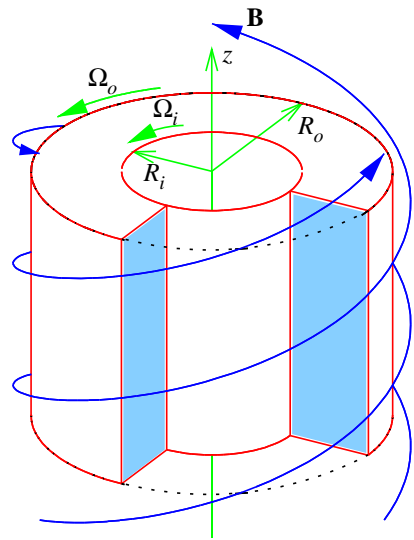} 
\par\end{centering}

\caption{\label{fig:sketch}Sketch to the formulation of the problem.}

\end{figure}

\section{\label{sec:prob-form}Problem formulation}

Consider an incompressible fluid of kinematic viscosity $\nu$ and
electrical conductivity $\sigma$ filling the gap between two infinite
concentric cylinders with inner radius $R_{i}$ and outer radius $R_{o}$
rotating with angular velocities $\Omega_{i}$ and $\Omega_{o}$,
respectively, in the presence of an externally imposed steady magnetic
field $\vec{B}_{0}=B_{\phi}\vec{e}_{\phi}+B_{z}\vec{e}_{z}$ with
axial and azimuthal components $B_{z}=B_{0}$ and $B_{\phi}=\beta B_{0}R_{i}/r$
in cylindrical coordinates $(r,\phi,z),$ where $\beta$ is a dimensionless
parameter characterizing the geometrical helicity of the field (Fig.
\ref{fig:sketch}). Further, we assume the magnetic field of the currents
induced by the fluid flow to be negligible relative to the imposed
field. This corresponds to the so-called inductionless approximation
holding for most of liquid-metal magnetohydrodynamics characterized
by small magnetic Reynolds numbers $\Rm=\mu_{0}\sigma v_{0}L\ll1,$
where $\mu_{0}$ is the magnetic permeability of vacuum and $v_{0}$
and $L$ are the characteristic velocity and length scale. The velocity
of fluid flow $\vec{v}$ is governed by the Navier-Stokes equation
with electromagnetic body force \begin{equation}
\frac{\partial\vec{v}}{\partial t}+(\vec{v}\cdot\vec{\nabla})\vec{v}=-\frac{1}{\rho}\vec{\nabla}p+\nu\vec{\nabla}^{2}\vec{v}+\frac{1}{\rho}\vec{j}\times\vec{B}_{0},\label{eq:N-S}\end{equation}
 where the induced current follows from Ohm's law for a moving medium
\begin{equation}
\vec{j}=\sigma\left(\vec{E}+\vec{v}\times\vec{B}_{0}\right).\label{eq:Ohm}\end{equation}
 In addition, we assume the characteristic time of velocity variation
to be much longer than the magnetic diffusion time $\tau_{0}\gg\tau_{m}=\mu_{0}\sigma L^{2}$
that leads to the quasi-stationary approximation, according to which
$\vec{\nabla}\times\vec{E}=0$ and $\vec{E}=-\vec{\nabla}\Phi,$ where
$\Phi$ is the electrostatic potential. Mass and charge conservations
imply $\vec{\nabla}\cdot\vec{v}=\vec{\nabla}\cdot\vec{j}=0.$

The problem admits a base state with a purely azimuthal velocity distribution
$\vec{v}_{0}(r)=\vec{e}_{\phi}v_{0}(r),$ where \[
v_{0}(r)=r\frac{\Omega_{o}R_{o}^{2}-\Omega_{i}R_{i}^{2}}{R_{o}^{2}-R_{i}^{2}}+\frac{1}{r}\frac{\Omega_{o}-\Omega_{i}}{R_{o}^{-2}-R_{i}^{-2}}.\]
 Note that the magnetic field does not affect the base flow because
it gives rise only to the electrostatic potential $\Phi_{0}(r)=B_{0}\int v_{0}(r)dr$
whose gradient compensates the induced electric field, so that there
is no current in the base state $(\vec{j}_{0}=0)$. However, a current
may appear in a perturbed state, \[
\left\{ \begin{array}{c}
\vec{v},p\\
\vec{j},\Phi\end{array}\right\} (\vec{r},t)=\left\{ \begin{array}{c}
\vec{v}_{0},p_{0}\\
\vec{j}_{0},\Phi_{0}\end{array}\right\} (r)+\left\{ \begin{array}{c}
\vec{v}_{1},p_{1}\\
\vec{j}_{1},\Phi_{1}\end{array}\right\} (\vec{r},t)\]
 where $\vec{v}_{1},$ $p_{1},$ $\vec{j}_{1},$ and $\Phi_{1}$ present
small-amplitude perturbations for which Eqs. (\ref{eq:N-S}) and (\ref{eq:Ohm})
after linearization take the form \begin{equation}
\frac{\partial\vec{v}_{1}}{\partial t}+(\vec{v}_{1}\cdot\vec{\nabla})\vec{v}_{0}+(\vec{v}_{0}\cdot\vec{\nabla})\vec{v}_{1}=-\frac{1}{\rho}\vec{\nabla}p_{1}+\nu\vec{\nabla}^{2}\vec{v}_{1}+\frac{1}{\rho}\vec{j}_{1}\times\vec{B}_{0},\label{eq:v1}\end{equation}
\begin{equation}
\vec{j}_{1}=\sigma\left(-\vec{\nabla}\Phi_{1}+\vec{v}_{1}\times\vec{B}_{0}\right).\label{eq:j1}\end{equation}
 In the following, we focus on axisymmetric perturbations which are
typically much more unstable than nonaxisymmetric ones \cite{Rued-ANN}.
For such perturbations the solenoidity constraints are satisfied by
meridional stream functions for fluid flow and electric current as
\[
\vec{v}=v\vec{e}_{\phi}+\vec{\nabla}\times(\psi\vec{e}_{\phi}),\qquad\vec{j}=j\vec{e}_{\phi}+\vec{\nabla}\times(h\vec{e}_{\phi}).\]
 Note that $h$ is the azimuthal component of the induced magnetic
field which is used subsequently instead of $\Phi$ for the description
of the induced current. Thus, we effectively retain the azimuthal
component of the induction equation to describe meridional components
of the induced current while the azimuthal current is explicitly related
to the radial velocity. The use of the electrostatic potential $\Phi,$
which provides an alternative mathematical formulation for the induced
currents in the inductionless approximation, would result in slightly
more complicated governing equations. In addition, for numerical purposes,
we introduce also the vorticity $\vec{\omega}=\omega\vec{e}_{\phi}+\vec{\nabla}\times(v\vec{e}_{\phi})=\vec{\nabla}\times\vec{v}$
as an auxiliary variable. The perturbation is sought in the normal
mode form \begin{equation}
\left\{ v_{1},\omega_{1,}\psi_{1},h_{1}\right\} (\vec{r},t)=\left\{ \hat{v},\hat{\omega},\hat{\psi},\hat{h}\right\} (r)e^{\gamma t+ikz},\label{eq:pert}\end{equation}
 where $\gamma$ is, in general, a complex growth rate and $k$ is
the axial wave number which is real for the conventional stability
analysis and complex for absolute instability. Henceforth, we proceed
to dimensionless variables by using $R_{i},$ $R_{i}^{2}/\nu,$ $R_{i}\Omega_{i},$
$B_{0},$ and $\sigma B_{0}R_{i}\Omega_{i}$ as the length, time,
velocity, magnetic field, and current scales, respectively. The nondimensionalized
governing equations then read as \begin{eqnarray}
\gamma\hat{v} & = & D_{k}\hat{v}+\RE ik(r^{2}\Omega)'r^{-1}\hat{\psi}+\Ha^{2}ik\hat{h},\label{eq:vhat}\\
\gamma\hat{\omega} & = & D_{k}\hat{\omega}+2\RE ik\Omega\hat{v}-\Ha^{2}ik(ik\hat{\psi}+2\beta r^{-2}\hat{h}),\label{eq:omghat}\\
0 & = & D_{k}\hat{\psi}+\hat{\omega},\label{eq:psihat}\\
0 & = & D_{k}\hat{h}+ik(\hat{v}-2\beta r^{-2}\hat{\psi}),\label{eq:hhat}\end{eqnarray}
 where $D_{k}f\equiv r^{-1}\left(rf'\right)'-(r^{-2}+k^{2})f$ and
the prime stands for $d/dr,$ $\RE=R_{i}^{2}\Omega_{i}/\nu$ and $\Ha=R_{i}B_{0}\sqrt{\sigma/(\rho\nu)}$
are Reynolds and Hartmann numbers, respectively, and \[
\Omega(r)=\frac{\lambda^{-2}-\mu+r^{-2}\left(\mu-1\right)}{\lambda^{-2}-1}\]
 is the dimensionless angular velocity of the base flow defined by
$\lambda=R_{o}/R_{i}$ and $\mu=\Omega_{o}/\Omega_{i}$. The boundary
conditions for the flow perturbation on the inner and outer cylinders
at $r=1$ and $r=\lambda,$ respectively, are $\hat{v}=\hat{\psi}=\hat{\psi}'=0.$
Boundary conditions for $\hat{h}$ on insulating and perfectly conducting
cylinders, respectively, are $\hat{h}=0$ and $(r\hat{h})'=0$ at
$r=1;\lambda.$

The governing equations (\ref{eq:vhat})--(\ref{eq:hhat}) for perturbation
amplitudes were discretized using a spectral collocation method on
a Chebyshev-Lobatto grid with a typical number of internal points
$N=32-96$. Auxiliary Dirichlet boundary conditions for $\hat{\omega}$
were introduced and then numerically eliminated to satisfy the no-slip
boundary conditions $\hat{\psi}'=0.$ The electric stream function
$\hat{h}$ was expressed in terms of $\hat{v}$ and $\hat{\psi}$
by solving Eq. (\ref{eq:hhat}) and then substituted in Eqs. (\ref{eq:vhat})
and (\ref{eq:omghat}) that eventually resulted in the $2N\times2N$
complex matrix eigenvalue problem which was solved by the LAPACK's
ZGEEV routine.

\section{Numerical results}

\subsection{\label{sub:Conv}Convective instability}

In this section, we consider the so-called convective instability
threshold supplied by the conventional linear stability analysis with
real wave numbers $k,$ as done in most of previous studies \cite{Hollerbach-Ruediger-2005,Stefani-etal,Priede-etal-2007}.
Note that at the convective instability threshold the system becomes
able to amplify certain perturbations which however might be not self-sustained
and, thus, experimentally unobservable without an external excitation.
The following results concern the radii ratio of outer to inner cylinder
$\lambda=2$ and we start with insulating cylinders which form the
side walls of the system.

\subsubsection{\label{sub:conv-insl}Insulating cylinders}

%
\begin{figure*}
\begin{centering}
\includegraphics[width=0.45\textwidth]{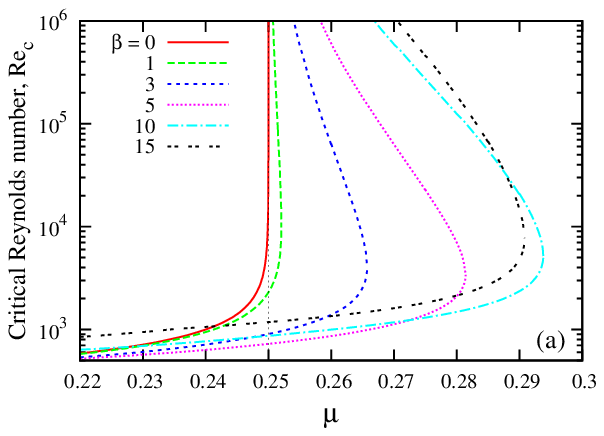}\includegraphics[width=0.45\textwidth]{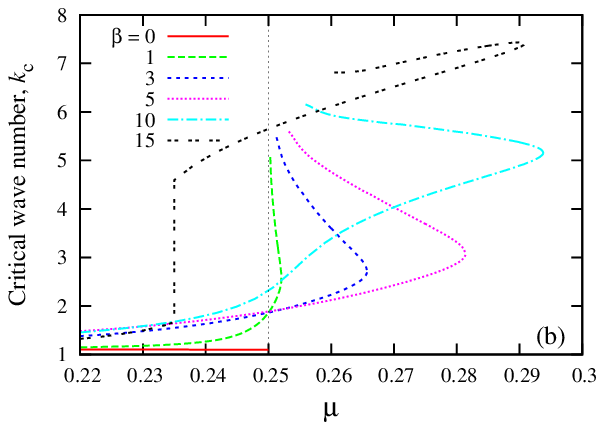}\\
 \includegraphics[width=0.45\textwidth]{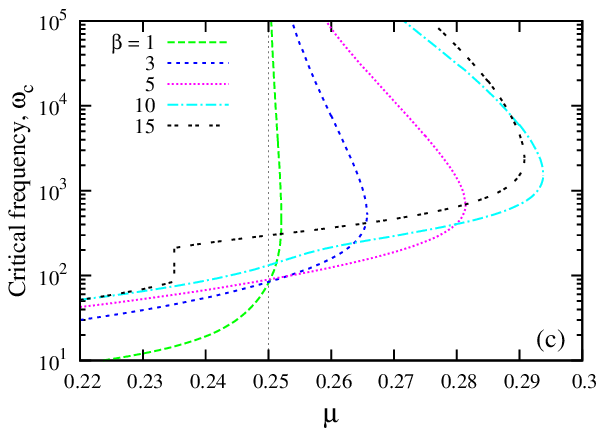} 
\par\end{centering}

\caption{\label{cap:Rec-mu}(a) Critical Reynolds number $\RE_{c}$, (b) wave
number $k_{c},$ and (c) frequency $\omega_{c}$ versus $\mu$ at
Hartmann number $\Ha=15$ and various helicities of the magnetic field
for insulating cylinders.}

\end{figure*}

The critical Reynolds number, wave number, and frequency are shown
in Fig. \ref{cap:Rec-mu} versus the angular velocity ratio $\mu$
of outer to inner cylinder for Hartmann number $\Ha=15$ and various
geometrical helicities $\beta.$ For $\beta=0$ corresponding to a
purely axial magnetic field, the critical Reynolds number tends to
infinity as $\mu$ approaches the Rayleigh line $\mu_{c}=\lambda^{-2}=0.25$
defined by $d\left(r^{2}\Omega\right)/dr=0.$ Thus, for $\beta=0,$
the range of instability is limited by the Rayleigh line, \textit{i.e.},
$\mu<\mu_{c},$ as in the purely hydrodynamic case. For helical magnetic
fields defined by $\beta\not=0,$ the instability extends well beyond
the Rayleigh line, as originally found by Hollerbach and R\"udiger
\cite{Hollerbach-Ruediger-2005}. Note that it is this extension
of the instability beyond its purely hydrodynamic limit, that for
ideal Taylor-Couette flow is defined by the Rayleigh line, which constitutes
the essence of the MRI. Comparing the stability curves presented in
Fig. \ref{cap:Rec-mu}(a) for fixed $\Ha$ to those of Hollerbach
and R\"udiger \cite{Hollerbach-Ruediger-2005}, which are presented
for $\Pm\not=0$ and variable $\Ha$ yielding minimal $\RE_{c},$
there are two differences to note. First, the range of instability
is limited by a certain $\mu_{\max},$ which depends on the helicity
$\beta$ and the Hartmann number, as shown in Fig. \ref{cap:muc-bt}(a).
Second, the destabilization beyond the Rayleigh line is effective
only in a limited range of Reynolds numbers bounded by an upper critical
value which tends to infinity as $\mu$ approaches the Rayleigh line
from the right.

%
\begin{figure*}[!]
\begin{centering}
\includegraphics[width=0.45\textwidth]{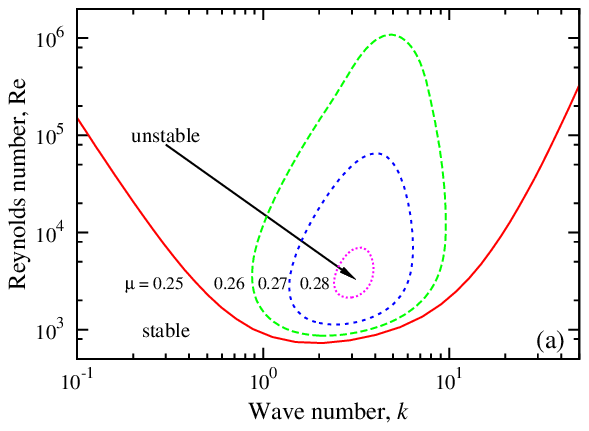}\includegraphics[width=0.45\textwidth]{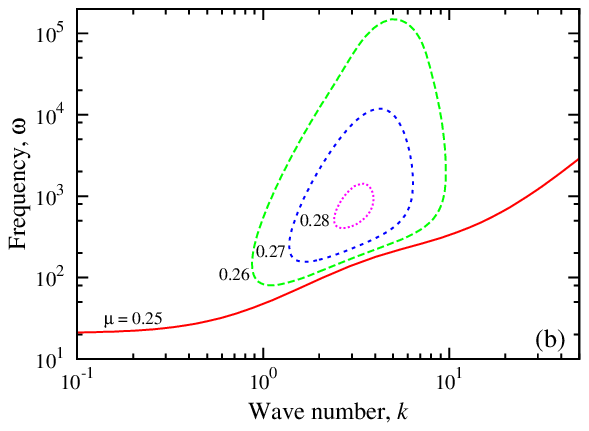} 
\par\end{centering}

\caption{\label{cap:re-wk}Neutral stability curves: (a) marginal Reynolds
number and (b) the frequency versus the wave number for $\beta=5$
and $\Ha=15$ at various ratios $\mu$ of angular velocities of the
outer to inner cylinder. }

\end{figure*}

The origin of the upper critical Reynolds number, by exceeding which
the flow becomes linearly stable again, is illustrated in Fig. \ref{cap:re-wk}
showing (a) the Reynolds number and (b) the corresponding frequency
of marginally stable modes versus their wave number $k$ for $\beta=5,$
$\Ha=15,$ and $\mu=0.27.$ As seen, the marginal stability curves
for $\mu$ beyond the Rayleigh line $(\mu>0.25)$ form closed loops
which collapse at $\mu=\mu_{\max}.$ Thus, unstable modes exist only
within limited ranges of wave and Reynolds numbers. Obviously, at
sufficiently large Reynolds numbers the flow becomes effectively non-magnetic
as inertia starts to dominate over the electromagnetic forces suppressing
the HMRI.

Note that the suppression of HMRI at high $\RE$ is related to the
negligibility of the induced magnetic field as long as $\Rm\ll1.$
In this case, the electric current is induced only by the velocity
perturbation crossing the imposed magnetic field. In the conventional
MRI conversely to the HMRI, also the induced field is relevant, which
crossed by the base flow induces an additional electric current. These
two effects are easily noticeable in the induction equation. Thus,
in the HMRI, the resulting electromagnetic force is not affected by
the base flow which is not the case for the conventional MRI, where
the electromagnetic force remains significant with respect to inertia
also at high $\RE.$

%
\begin{figure*}
\begin{centering}
\includegraphics[width=0.45\textwidth]{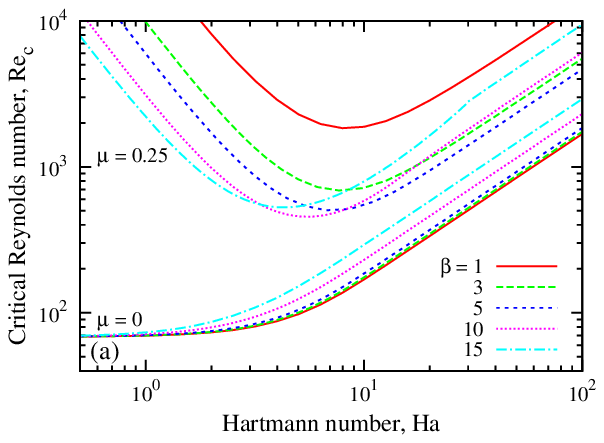}\includegraphics[width=0.45\textwidth]{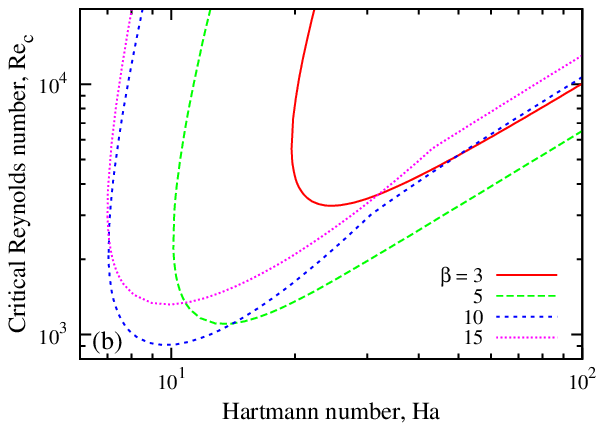}\\
 \includegraphics[width=0.45\textwidth]{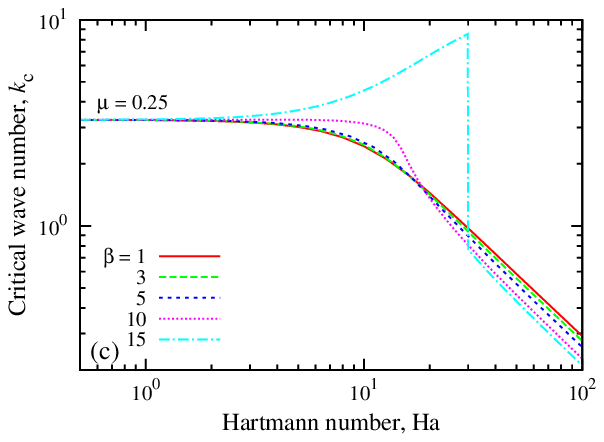}\includegraphics[width=0.45\textwidth]{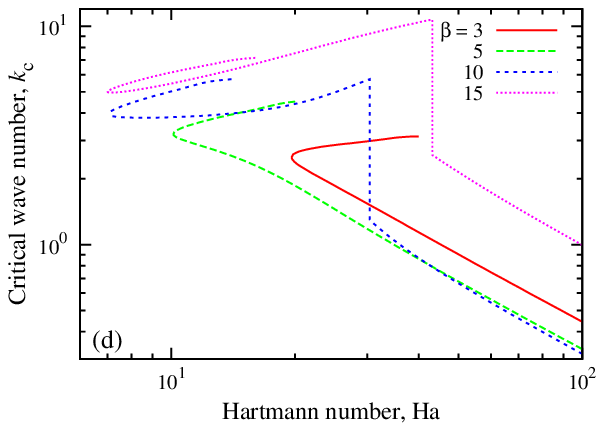}\\
 \includegraphics[width=0.45\textwidth]{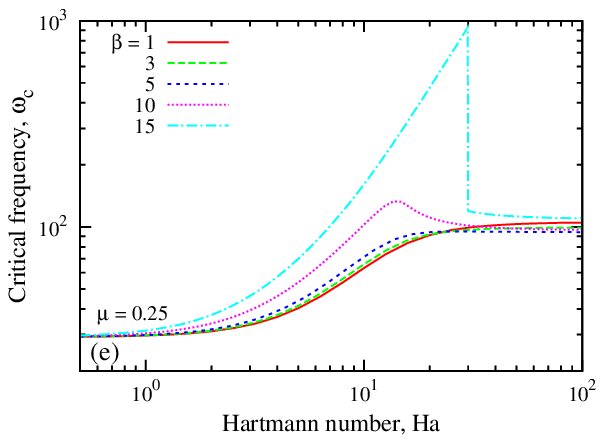}\includegraphics[width=0.45\textwidth]{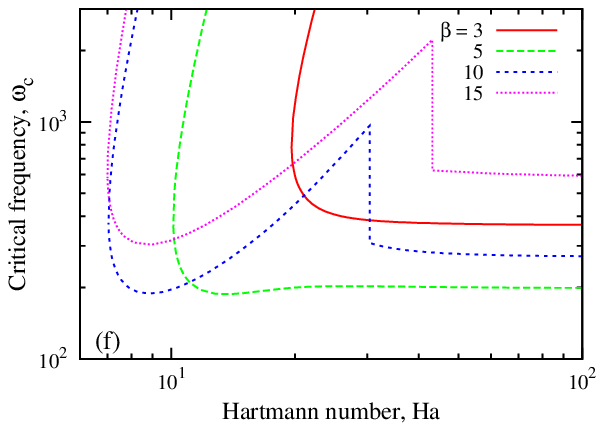} 
\par\end{centering}

\caption{\label{cap:Rec-Ha} {[}(a) and (b)] Critical Reynolds number $\RE_{c}$,
{[}(c) and (d)] wave number $k_{c}$, and {[}(e) and (f)] frequency
$\omega_{c}$ versus the Hartmann number for (a) $\mu=0$, {[}(a),
(c), and (e)] $0.25,$ and {[}(b), (d), and (f)] $0.27$ at various
helicities $\beta$ and insulating cylinders.}

\end{figure*}

As seen in Figs. \ref{cap:Rec-Ha}(a) and \ref{cap:Rec-Ha}(b), the
critical Reynolds number can vary with the Hartmann number in three
different ways depending on $\mu.$ For $\mu=0,$ the critical Reynolds
number is bounded at $\Ha=0$ because a purely hydrodynamic instability
is possible before the Rayleigh line. Numerical results evidence that
the increase in the Hartmann number results in the growth of the critical
Reynolds number with asymptotics $\sim\Ha.$ For $\mu=0.25,$ which
lies exactly on the Rayleigh line, the flow is hydrodynamically stable
without the magnetic field. Thus, in this case, the critical Reynolds
number increases as $\sim\Ha^{-2}$ as $\Ha\rightarrow0$ because
there is no finite value of the critical Reynolds number without the
magnetic field. The corresponding critical wave number tends to a
finite value independent of $\beta.$ With increase in the Hartmann
number, the critical Reynolds number attains a minimum at $\Ha\sim10$
and starts to grow at larger Hartmann numbers similarly to the previous
case. The corresponding critical wave number decreases asymptotically
as $\sim\Ha^{-1}$ that means a critical wavelength increasing directly
with the magnetic field strength. The critical frequency plotted in
Fig. \ref{cap:Rec-Ha} changes from a constant value of 30 at small
$\Ha$ to another nearly constant value of about $100$ slightly varying
with $\beta$ at large $\Ha.$ At large helicities $(\beta=15),$
another relatively short-wave instability mode dominates up to a Hartmann
number $\Ha\approx30,$ where the most unstable mode switches back
to the long-wave one which is characteristic for smaller helicities.
Transition to this large-$\beta$ mode is also obvious in Fig. \ref{cap:Rec-mu}
for $\beta=15$ at $\mu\approx0.235.$ As seen in Fig. \ref{cap:Rec-Ha}(b)
for $\mu=0.27$, which is beyond the Rayleigh line, there is no instability
as $\Ha\rightarrow0.$ Consequently, a finite minimal value of $\Ha$
depending on $\beta$ is necessary in this case. Moreover, the instability
is limited by the upper branch of the critical Reynolds number discussed
above which merges with the lower branch at the minimum of the Hartmann
number for the given helicity $\beta.$

%
\begin{figure*}
\begin{centering}
\includegraphics[width=0.45\textwidth]{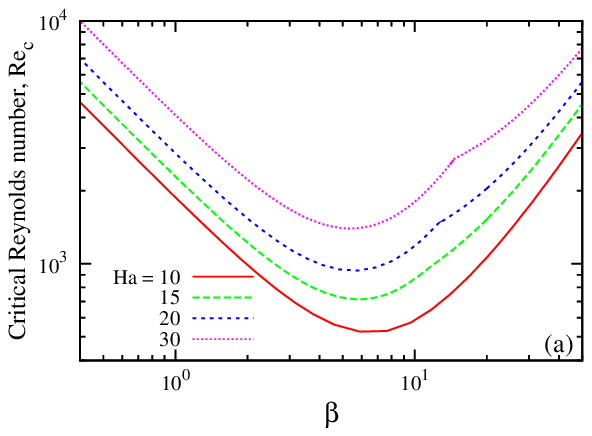}\includegraphics[width=0.45\textwidth]{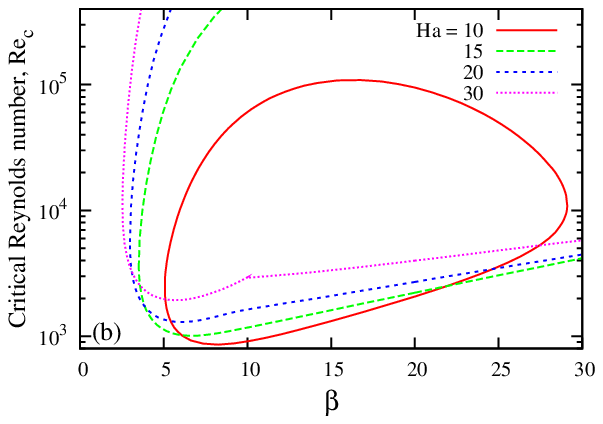}\\
 \includegraphics[width=0.45\textwidth]{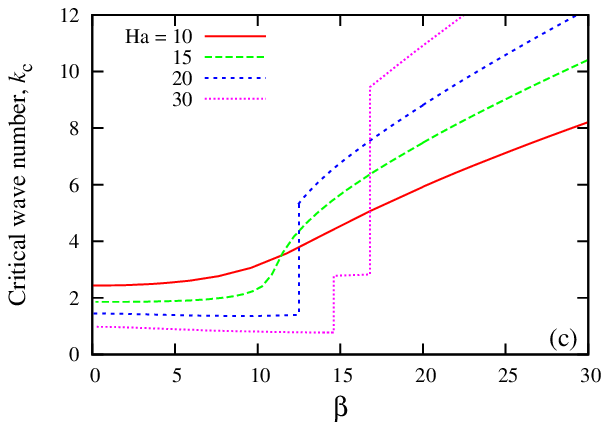}\includegraphics[width=0.45\textwidth]{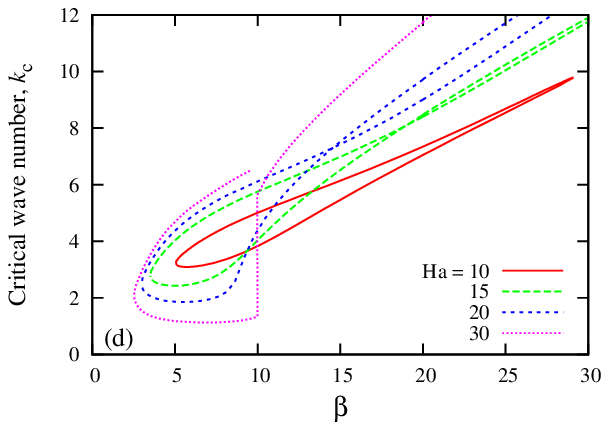}\\
 \includegraphics[width=0.45\textwidth]{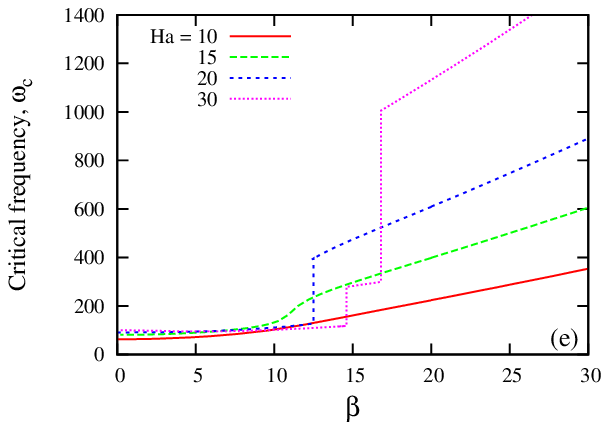}\includegraphics[width=0.45\textwidth]{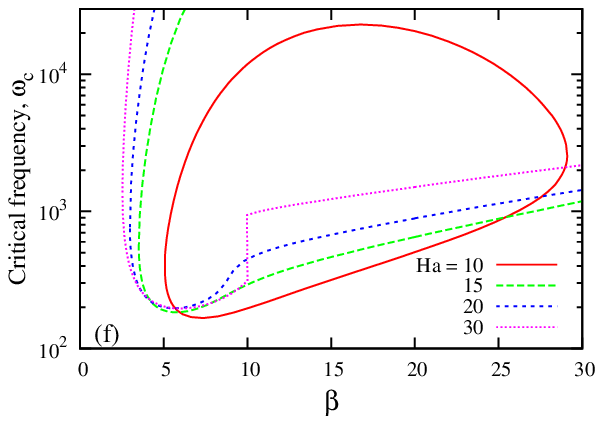} 
\par\end{centering}

\caption{\label{cap:Rec-bt} {[}(a) and (b)] Critical Reynolds number $\RE_{c},$
{[}(c) and (d)] wave number $k_{c},$ and {[}(e) and (f)] frequency
$\omega_{c}$ versus $\beta$ for {[}(a), (c), and (e)] $\mu=0.25$
and {[}(b), (d), and (f)] $\mu=0.27$ at various Hartmann numbers. }

\end{figure*}

The variation of the critical Reynolds number with the helicity $\beta$
shown in Fig. \ref{cap:Rec-bt}(a) for $\mu=0.25=\mu_{c}$ lying exactly
on the Rayleigh line differs considerably from the other case with
$\mu=0.27>\mu_{c}$ {[}see Fig. \ref{cap:Rec-bt}b]. In the first
case, the flow can be destabilized by the magnetic field of however
small helicity $\beta\rightarrow0$ that results in the critical Reynolds
number increasing as $\sim1/\beta.$ For $\mu>\mu_{c},$ a certain
minimal helicity depending on the Hartmann number is needed. Moreover,
in this case, there is also an upper critical Reynolds number. In
both cases, there is some optimal $\beta\approx5-8$ at which the
lower critical Reynolds attains a minimum. Further increase in $\beta$
results in the growth of the critical Reynolds number with a significantly
different asymptotic behavior in both considered cases. For $\mu>\mu_{c},$
there is a maximal $\beta$ depending on the Hartmann number at which
the upper and lower branches of the critical Reynolds number merge
together and the instability disappears whereas there seems to be
no such merging point at any finite $\beta$ when $\mu=\mu_{c}.$
The critical wave number plotted in Figs. \ref{cap:Rec-bt}(c) and
\ref{cap:Rec-bt}(d) is seen to increase with $\beta$ with some jumps
at larger $\Ha$ as discussed above.

\subsubsection{Perfectly conducting cylinders}

%
\begin{figure*}
\begin{centering}
\includegraphics[width=0.45\textwidth]{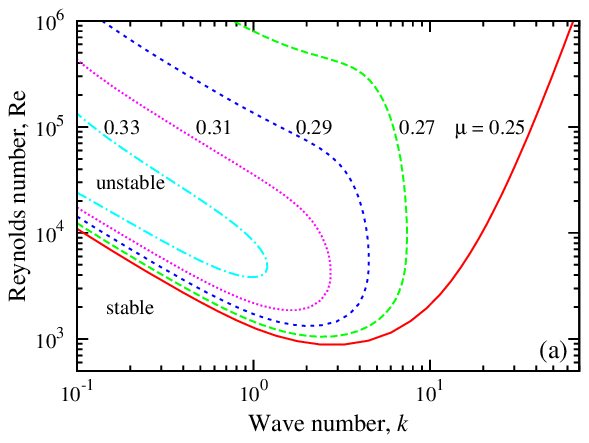}\includegraphics[width=0.45\textwidth]{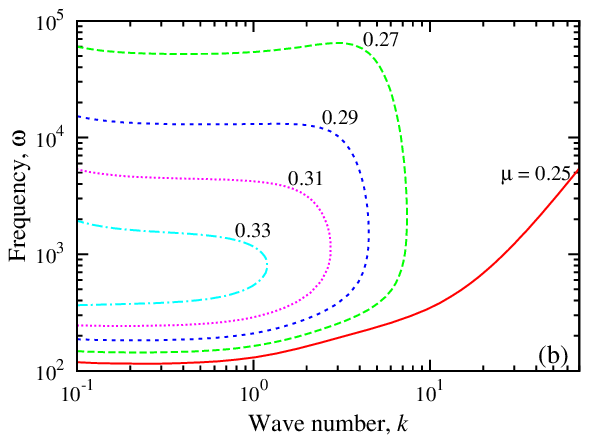} 
\par\end{centering}

\caption{\label{cap:re-wk-c}Neutral stability curves: (a) marginal Reynolds
number and (b) the frequency versus the wave number for $\beta=5$
and $\Ha=15$ at various ratios $\mu$ of angular velocities of the
outer to inner cylinder. }

\end{figure*}

For perfectly conducting cylinders, the marginal stability curves
shown in Fig. \ref{cap:re-wk-c} differ considerably from those for
insulating walls (see Fig. \ref{cap:re-wk}). Although in both cases
beyond the Rayleigh line large wave numbers $(k\gg1)$ are always
stable, the range of instability for perfectly conducting cylinders
at moderate $\beta\lesssim10$ extends to arbitrary small wave numbers
$k\rightarrow0$ whereas for insulating cylinders it is limited to
sufficiently large $k$. As in the insulating case, for each unstable
mode there is not only the lower but also the upper marginal Reynolds
number both increasing as $\sim1/k$ toward small $k.$ Thus, the
increase in the Reynolds number results in the shift of instability
to smaller wave numbers, \textit{i.e.}, longer waves. As a result,
there is no upper critical Reynolds number for moderate $\beta\lesssim10$
when both cylinders are perfectly conducting.

%
\begin{figure*}
\begin{centering}
\includegraphics[width=0.45\textwidth]{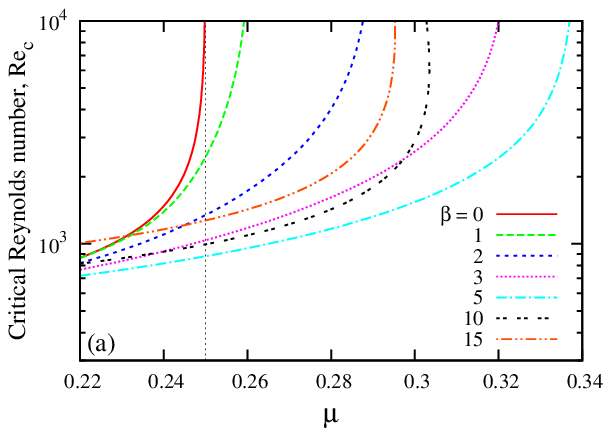}\includegraphics[width=0.45\textwidth]{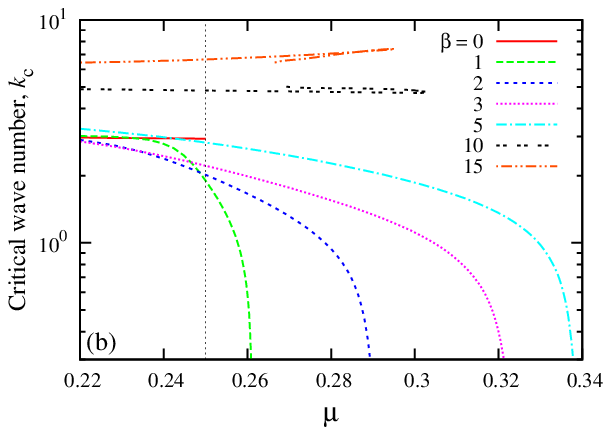}\\
 \includegraphics[width=0.45\textwidth]{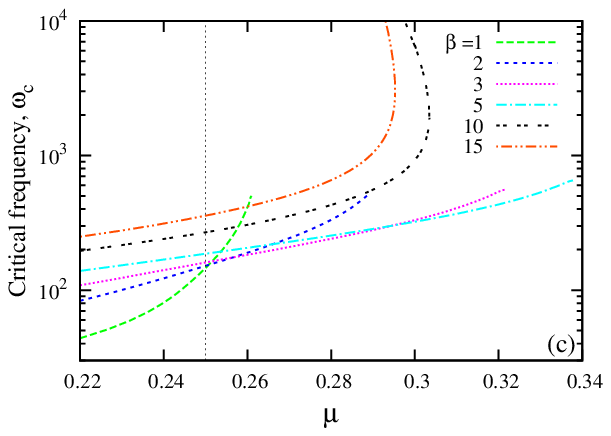} 
\par\end{centering}

\caption{\label{cap:Rec-mu-c} (a) Critical Reynolds number $\RE_{c},$ (b)
wave number $k_{c},$ and (c) frequency $\omega_{c}$ versus $\mu$
for Hartmann numbers $\Ha=15$ and various magnetic field helicities
(perfectly conducting cylinders). }

\end{figure*}

Consequently, as seen in Figs. \ref{cap:Rec-mu-c}(a) and \ref{cap:Rec-mu-c}(b),
the critical Reynolds number becomes very large while the critical
wave number tends to zero as $\mu$ approaches some critical $\mu_{\max}$
which varies with $\beta.$ The critical frequency $\omega_{c}$ shown
in Fig. \ref{cap:Rec-mu-c}(c) tends, respectively, to some finite
value. This behavior changes at larger $\beta$ becoming similar to
that for insulating cylinders. As seen in Fig. \ref{cap:Rec-mu-c}(a),
for $\beta\gtrsim10,$ the curves of the critical Reynolds number
start to bend back at $\mu_{\max}$ toward smaller $\mu$ rather than
tend to infinity. The corresponding critical wave numbers remain finite
whereas the critical frequency increases with the Reynolds number
{[}see Figs. \ref{cap:Rec-mu-c}b and \ref{cap:Rec-mu-c}c]. At intermediate
$\beta$ the limiting value of $\mu_{\max},$ up to which the instability
extends beyond the Rayleigh line, is seen in Fig. \ref{cap:muc-bt}(a)
to attain a maximum which is considerably larger than that for insulating
walls. At larger $\beta$, the limiting values $\mu_{\max}$ decrease
approaching those for insulating walls.

%
\begin{figure*}
\begin{centering}
\includegraphics[width=0.45\textwidth]{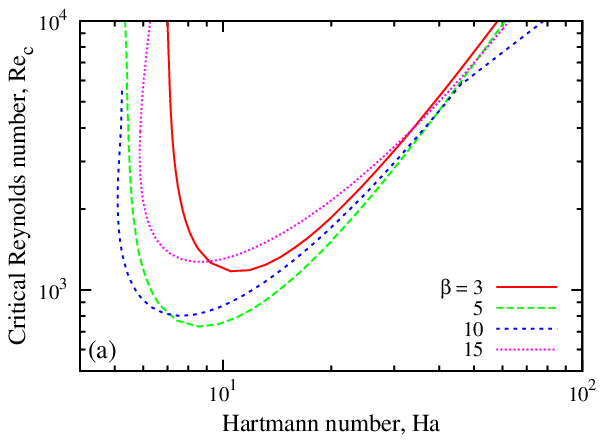}\includegraphics[width=0.45\textwidth]{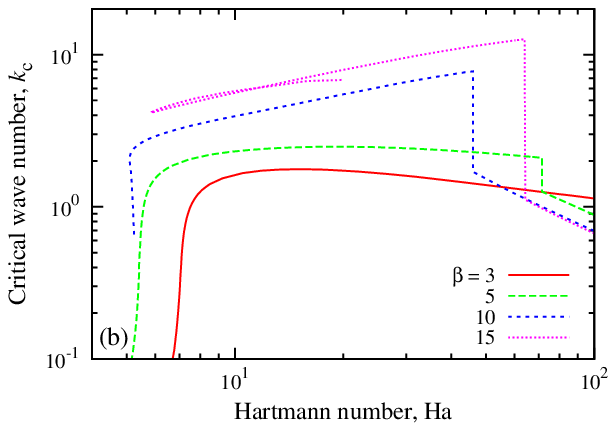}\\
 \includegraphics[width=0.45\textwidth]{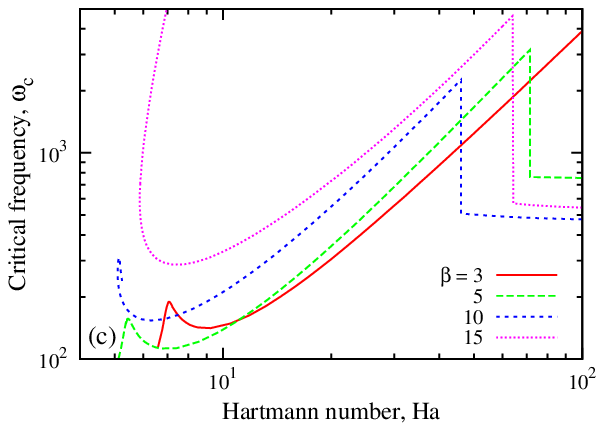} 
\par\end{centering}

\caption{\label{cap:Rec-Ha-c}(a) Critical Reynolds number $\RE_{c},$ (b)
wave number $k_{c},$ and (c) frequency $\omega_{c}$ versus the Hartmann
number for $\mu=0.27$ at various helicities $\beta$ and perfectly
conducting cylinders.}

\end{figure*}

The dependence of the critical Reynolds number on the Hartmann number
plotted in Fig. \ref{cap:Rec-Ha-c}(a) at various helicities is similar
to that for insulating cylinders. First, $\RE_{c}$ attains a minimum
at $\Ha=7-10$ and a finite minimal value of the Hartmann number is
required for instability when $\mu>\mu_{c}.$ For moderate $\beta\lesssim10,$
in contrast to the insulating case, $\RE_{c}$ and $k_{c}$ tend to
infinity and zero, respectively, as the Hartmann number approaches
this minimal value which depends on $\beta.$ For $\beta\gtrsim10,$
the critical $\RE_{c}$ has an upper branch which merges with the
lower one at the minimal value of $\Ha$ as in the case of insulating
cylinders. At sufficiently large Hartmann numbers, the instability
is seen to switch to a long-wave mode with the critical wave numbers
and Reynolds numbers varying asymptotically as $\sim\Ha^{-1}$ and
$\sim\Ha,$ respectively.

\subsection{\label{sub:Abs-inst}Absolute instability}

In this section, we turn to the absolute instability for which the
wave number $k$ is in general a complex quantity with real and imaginary
parts $k_{r}$ and $k_{i},$ respectively \cite{Lifshitz-Pitaevskii-81,Schmid-Henningson}.
It is important to realize that the convective instability threshold
considered above is not sufficient for the development of a self-sustained
instability unless the system is mirror symmetric along the direction
of propagation which, however, is not the case when the magnetic field
is helical. The convective instability just ensures the ability of
the system to amplify external perturbations excited with the critical
frequency. From the mathematical point of view, the problem is that
in an axially bounded system the perturbation has to meet certain
boundary conditions at two end walls that, however, can not be accomplished
by a single Fourier mode. When the critical Fourier mode is replaced
with a corresponding wave packet of a limited spatial extension, one
finds such a perturbation to grow only in the frame of reference traveling
with its group velocity while it decays asymptotically in any other
frame of reference including the laboratory one which is at rest.
The growth of a perturbation in the laboratory frame of reference
is ensured by the absolute instability threshold, at which the group
velocity of the wave packet becomes zero. Thus, formally, the absolute
instability requires one more condition to be satisfied, \textit{i.e.},
zero group velocity, by means of an additional free parameter---the
imaginary part of the wave number. Note that although the group velocity
in non-conservative media is, in general, a complex quantity, for
the most dangerous perturbations satisfying $\partial\gamma_{r}/\partial k_{r}=0,$
it is real and, thus, coincides with its common definition.

Alternatively, the absolute instability may be regarded as an asymptotic
case of the global instability when the axial extension of the system
becomes very large \cite{Landau-87,Lifshitz-Pitaevskii-81,Kulikovskii-66}.
The basic idea is that for a convectively unstable perturbation to
become self-sustained a feedback mechanism is needed which could transfer
a part of the growing perturbation as it leaves the system back to
its origin. Such a feedback can be provided by the reflections of
perturbation from the end walls or, generally, by the end regions
where the base state becomes axially non-uniform. If the base state
is both stationary and axially uniform, the coefficients of the linearized
perturbation equations do not depend on time and on the axial coordinate,
respectively. Then, as for linear differential equations with constant
coefficients, the particular solution for the perturbation varies
exponentially in both time and axial coordinate as supposed by Eq.
(\ref{eq:pert}) where both the growth rate $\gamma$ and the wave
number $k$ may be in general complex. At the end walls, where the
base state is no longer axially invariant, the particular solutions
with different wave numbers become linearly coupled while their time
variation remains unaffected as long as the base state is stationary.
Thus, the reflection of a perturbation by the end wall in general
couples modes with different wave numbers but with the same $\gamma.$
Sufficiently far away from the end walls the reflected perturbation
is dominated by the mode with the imaginary part of the wave number
$k_{i}$ corresponding to either the largest growth or lowest decay
rate along the axis. Consequently, sufficiently away from the end
walls a global mode is expected to consist of two such waves coupled
by reflections from the opposite end walls. Taking into account that
the amplitude of the reflected wave is proportional to that of the
incident wave, it is easy to find that in a sufficiently extended
system both waves must have the same imaginary part $k_{i}$ of the
wave number whereas the real parts may be different. Additionally,
for two such waves to be coupled by reflections from the end walls,
they have to propagate in opposite directions.

\subsubsection{Insulating cylinders}

%
\begin{figure*}
\begin{centering}
\includegraphics[width=0.45\textwidth]{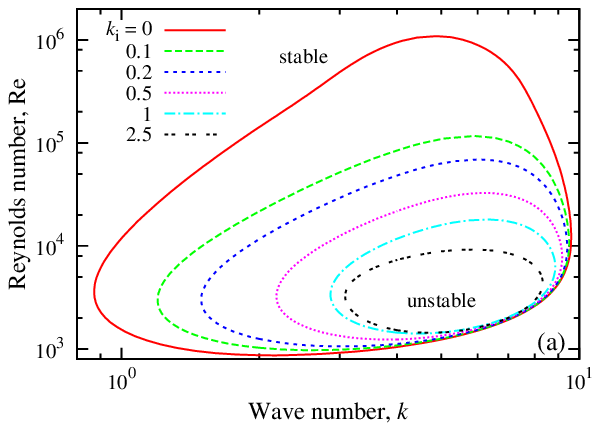}\includegraphics[width=0.45\textwidth]{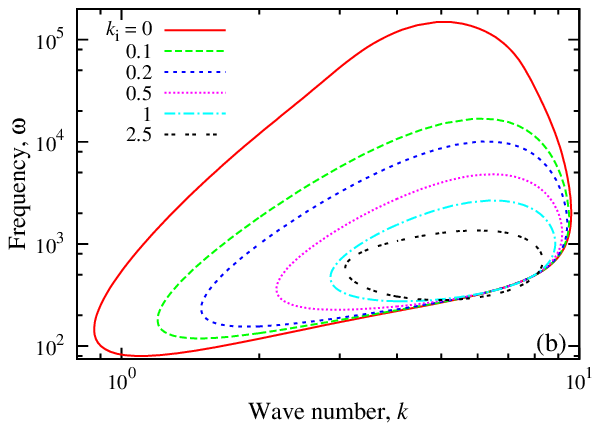}\\
 \includegraphics[width=0.45\textwidth]{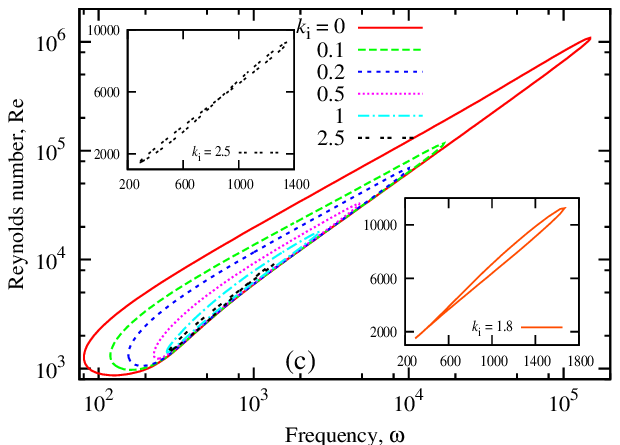} 
\par\end{centering}

\caption{\label{cap:re-wki}Neutral stability curves: (a) marginal Reynolds
number and (b) the frequency versus the wave number, and (c) the Reynolds
number versus the frequency for $\mu=0.26,$ $\beta=5,$ and $\Ha=15$
at various imaginary parts of wave number $k_{i}.$}

\end{figure*}

We search for such a pair of modes by considering the conventional
neutral stability curves at various $k_{i}.$ As seen in Figs. \ref{cap:re-wki}(a)
and (b), the increase in $k_{i},$ on one hand, results in the reduction
in the wave number range admitting such neutrally stable modes. On
the other hand, the lower branch of marginal $\RE$ and the corresponding
frequency first increase with $k_{i}$ in the whole wave number range
and then start to decrease at larger $k_{r}$ when $k_{i}$ becomes
sufficiently large $(k_{i}\gtrsim1.8).$ However, more important information
is obtained by plotting the marginal $\RE$ and frequency from the
previous curves against each other as in Fig. \ref{cap:re-wki}(c).
First, similarly to the previous curves, these ones also form closed
loops that shrink as $k_{i}$ is increased. It is important to notice
that at sufficiently large $k_{i}$ these loops start to intersect
themselves in some point as shown in the inset at the top of Fig.
\ref{cap:re-wki}(c). This self-intersection, which is of primary
importance here, occurs only in the limited range of positive $k_{i}.$
The point of intersection means that at the given Reynolds number
there are two modes with the same frequency and the same imaginary
but possibly different real parts of the wave number. As discussed
above, two such modes could be coupled by reflections from the end
walls and thus form a neutrally stable global mode in an axially bounded
system provided that they propagate in opposite directions \cite{Lifshitz-Pitaevskii-81,Kulikovskii-66}.
To determine the direction of propagation we use a local criterion
\cite{Priede-Gerbeth-97} which we showed to be equivalent to the
Briggs pinching criterion \cite{Briggs-1964} for the upper instability
branch at the given $k_{i}.$ Namely, the direction of propagation
of both intersecting branches can be deduced from their variation
with $k_{i}.$ If upon a small variation of $k_{i}$ one branch rises
to higher $\RE$ while the other descends to lower $\RE$, which is
the case here, it can be shown that both intersecting branches correspond
to oppositely propagating modes. The lowest possible Reynolds number
admitting two such modes is attained when the loop below the intersection
point collapses to a cusp as seen in the inset at the bottom of Fig.
\ref{cap:re-wki}(c) for $k_{i}=1.8.$ The cusp is formed as both
intersection points of the loop merge together. It means that at the
cusp point not only the imaginary but also the real parts of both
wave numbers become equal. This point corresponds to the absolute
instability at which the length of the wave packet of the global mode
formed by two waves with merging wave numbers tends to infinity. Further
we focus on this absolute instability which, in contrast to the convective
one considered above, can be self-sustained in a sufficiently extended
system. Note that the approach outlined above to find the absolute
instability is an extension of the well-known cusp map for the complex
frequency plane to the ($\RE$-$\omega$) plane \cite{Schmid-Henningson,Kupfer_87}
by using the neutral stability condition $\lambda_{r}(\RE)=0$ which
maps the real part of the growth rate $\lambda_{r}$ to the marginal
Reynolds number.

%
\begin{figure*}
\begin{centering}
\includegraphics[width=0.45\textwidth]{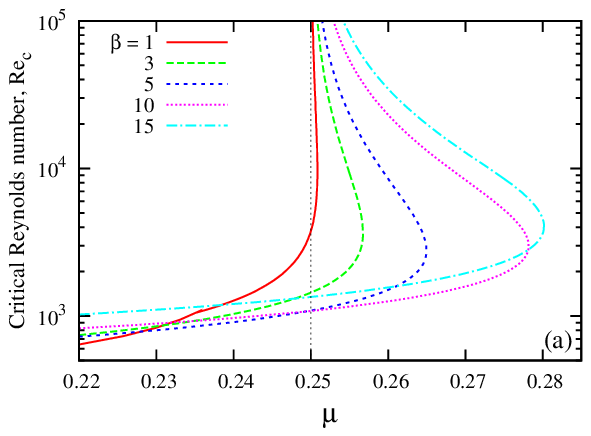}\includegraphics[width=0.45\textwidth]{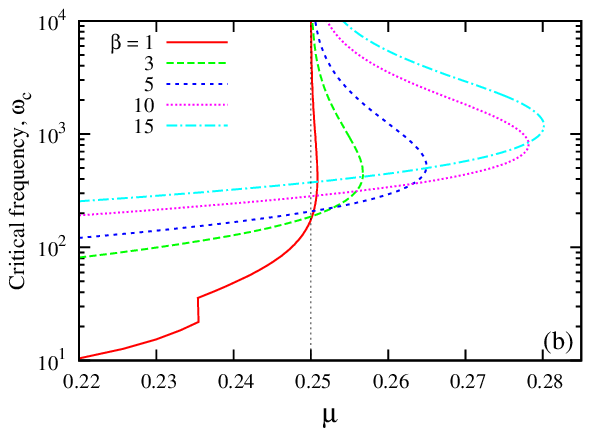}\\
 \includegraphics[width=0.45\textwidth]{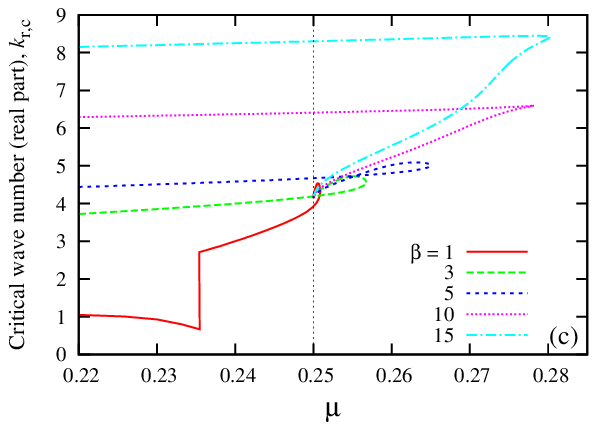}\includegraphics[width=0.45\textwidth]{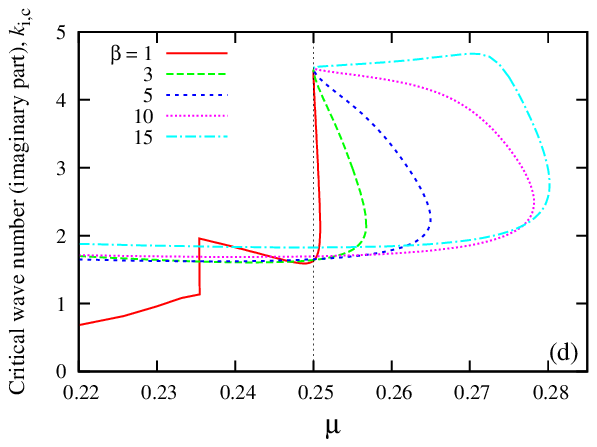} 
\par\end{centering}

\caption{\label{cap:Reg-mu}(a) Critical Reynolds number $\RE_{c},$ (b) frequency
$\omega_{c},$ (c) real and (d) imaginary parts of wave number $k_{r,c}$
and $k_{i,c}$ versus $\mu$ at $\Ha=15$ and various magnetic field
helicities for insulating cylinders. }

\end{figure*}

The critical Reynolds number, frequency, and the critical complex
wave number for the absolute instability threshold is plotted in Fig.
\ref{cap:Reg-mu} versus $\mu$ at $\Ha=15$ and various helicities
of the magnetic field. Comparison with the corresponding convective
instability, the critical parameters of which are plotted in Fig.
\ref{cap:Rec-mu}, shows that before the Rayleigh line the threshold
of absolute instability is only slightly above the convective one.
The difference between both thresholds becomes significant at the
Rayleigh line. Although the absolute instability extends beyond the
Rayleigh line when the magnetic field is helical, the range of extension
is noticeably shorter than that of the convective instability (see
Fig. \ref{cap:muc-bt}). Moreover, the upper critical Reynolds number
for the absolute instability is considerably lower than that of the
convective one. Although the difference between the critical Reynolds
numbers for the absolute and convective instabilities is insignificant
before the Rayleigh line, the critical wave numbers for the absolute
instability shown in Fig. \ref{cap:Reg-mu}(c) are considerably larger
than those for the convective instability {[}see Fig. \ref{cap:Rec-mu}b].
This difference increases with $\beta$ that results in the rise of
the critical wave number for absolute instability while the increase
in the corresponding quantity for the convective instability threshold
is insignificant. In contrast to this, the imaginary part of the critical
wave number for the lower instability branch $k_{i}\approx1.8$ is
almost invariable with both $\beta$ and $\mu$ except for $\beta=1,$
where a jump of the instability to a larger wave number takes place
at $\mu\approx0.235.$ Note that positive $k_{i}$ corresponds to
the amplitude of the critical perturbation growing axially downward
which is an additional feature predicted by the absolute instability.
Beyond the Rayleigh line the absolute instability similarly to the
convective one is effective only in a limited range of Reynolds numbers
which is bounded from above by the upper critical branch tending to
the Rayleigh line from the right as the Reynolds number increases.
The critical complex wave number is seen in Figs. \ref{cap:Reg-mu}(c)
and \ref{cap:Reg-mu}(d) to tend to a certain limiting value independent
of $\beta.$

\subsubsection{Perfectly conducting cylinders}

%
\begin{figure*}
\begin{centering}
\includegraphics[width=0.45\textwidth]{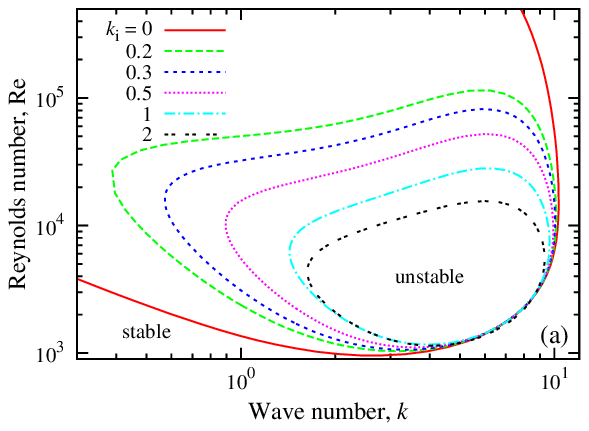}\includegraphics[width=0.45\textwidth]{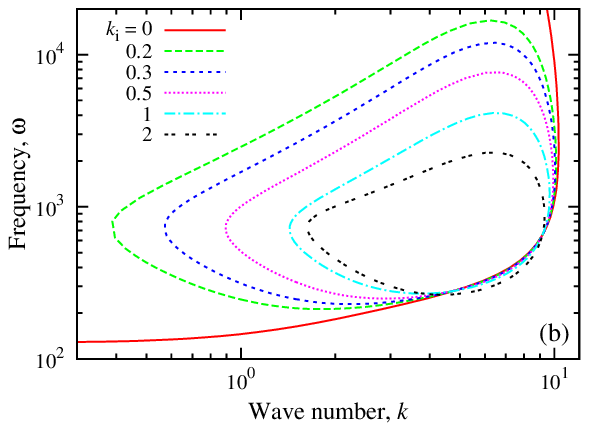}\\
 \includegraphics[width=0.45\textwidth]{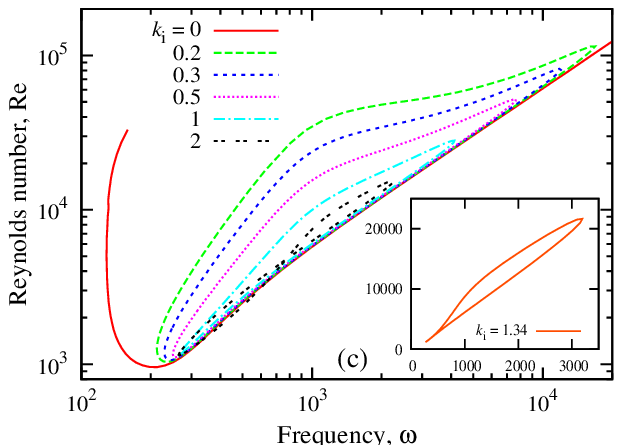} 
\par\end{centering}

\caption{\label{cap:re-wki-c} Neutral stability curves: (a) marginal Reynolds
number, (b) the frequency versus the wave number, and (c) the Reynolds
number versus the frequency for $\mu=0.26,$ $\beta=5$ and $\Ha=15$
at various imaginary parts of wave number $k_{i}$ for perfectly conducting
cylinders.}

\end{figure*}

The neutral stability curves for perfectly conducting cylinders plotted
in Fig. \ref{cap:re-wki-c} are seen to start forming closed loops
when $k_{i}>0$ becoming similar to the corresponding curves for insulating
cylinders shown in Fig. \ref{cap:re-wki}. In a certain range of $k_{i}$
the curves of the critical Reynolds number plotted against the frequency
in Fig. \ref{cap:re-wki-c}(c) intersect themselves that implies the
existence of two neutrally stable modes with the same Reynolds number,
frequency, and imaginary part of the wave number but different real
parts of the wave number. As discussed above, two such modes can be
coupled by reflections from the end walls and thus form a neutrally
stable small-amplitude global mode in the system of a large but finite
axial extension provided that those modes propagate in opposite directions
that is implied by the variation of the marginal Reynolds number upon
a small variation of $k_{i}.$

%
\begin{figure*}
\begin{centering}
\includegraphics[width=0.45\textwidth]{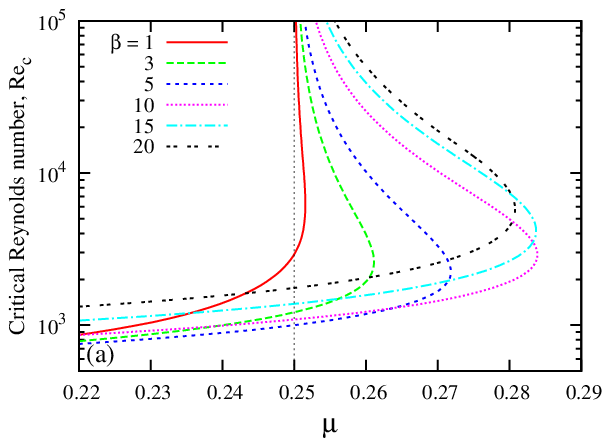}\includegraphics[width=0.45\textwidth]{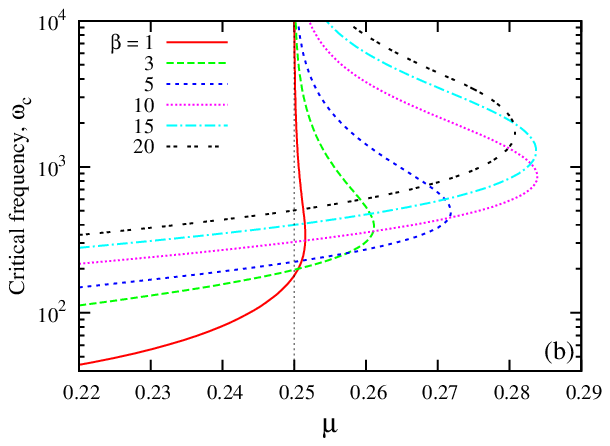}\\
 \includegraphics[width=0.45\textwidth]{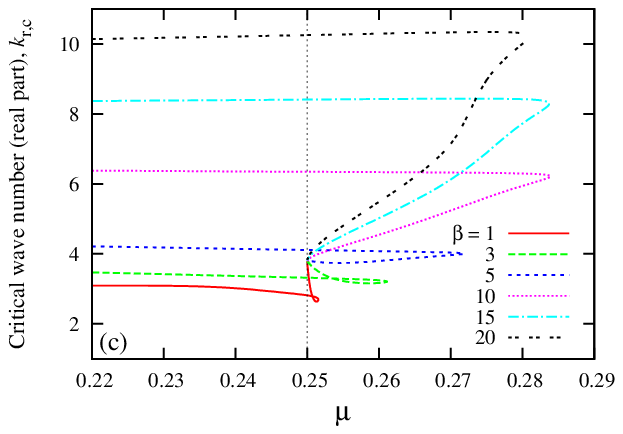}\includegraphics[width=0.45\textwidth]{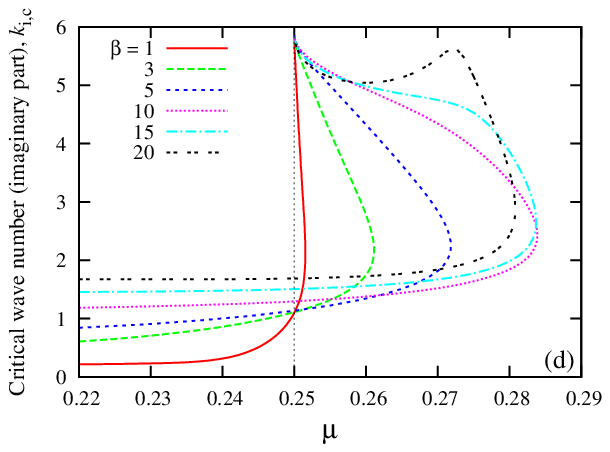} 
\par\end{centering}

\caption{\label{cap:Reg-mu-c} (a) Critical Reynolds number $\RE_{c},$ (b)
frequency $\omega_{c},$ (c) real and (d) imaginary parts of wave
number $k_{r,c}$ and $k_{i,c}$ versus $\mu$ for $\Ha=15$ and various
magnetic field helicities $\beta$ for perfectly conducting cylinders. }

\end{figure*}

\begin{figure*}
\begin{centering}
\includegraphics[width=0.45\textwidth]{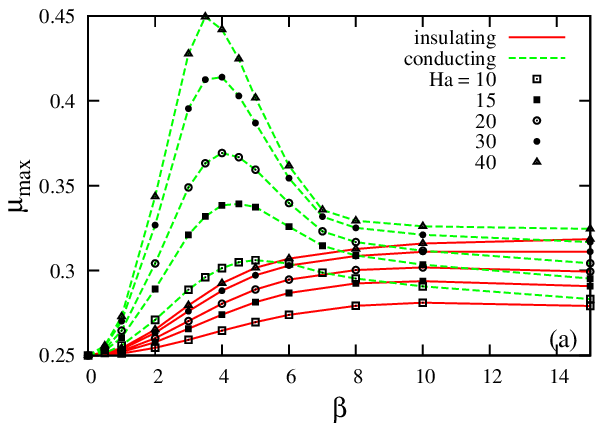}\includegraphics[width=0.45\textwidth]{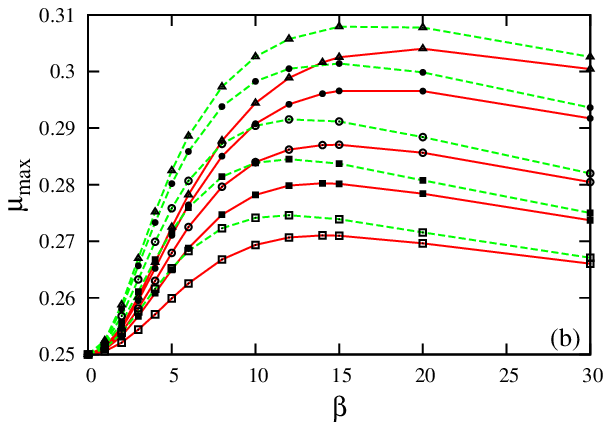} 
\par\end{centering}

\caption{\label{cap:muc-bt} Maximal value of $\mu$ versus the magnetic field
helicity $\beta$ for (a) convective and (b) absolute instability
thresholds at various Hartmann numbers for both insulating and perfectly
conducting cylinders with $\lambda=2.$ }

\end{figure*}

For perfectly conducting cylinders, the critical Reynolds number,
frequency, and complex wave number for the absolute instability threshold
plotted in Fig. \ref{cap:Reg-mu-c} versus $\mu$ for $\Ha=15$ and
various helicities differ significantly from the corresponding critical
parameters for the convective instability threshold (see Fig. \ref{cap:Rec-mu-c}).
First, the range of extension of the absolute instability beyond the
Rayleigh line is much shorter than that of the convective instability.
Note that in contrast to the convective instability there is no significant
difference with respect to the extension of the absolute instability
beyond the Rayleigh line between insulating and perfectly conducting
cylinders (see Fig. \ref{cap:muc-bt}). Second, beyond the Rayleigh
line, similarly to insulating cylinders, for all $\beta$ the range
of unstable Reynolds numbers is bounded from above by the upper critical
branches which approach the Rayleigh line from the right as the upper
critical Reynolds number tends to infinity. Similarly to the insulating
cylinders, the corresponding critical complex wave number tends to
a certain asymptotic value independent of $\beta$ {[}see Figs. \ref{cap:Reg-mu-c}c
and \ref{cap:Reg-mu-c}d]. Third, beyond the Rayleigh line the critical
wave numbers for the absolute instability are noticeably greater than
those for the convective instability, especially for $\beta\lesssim10$
when the critical wave numbers for the convective instability tend
to zero {[}see Fig. \ref{cap:Rec-mu-c}b].

%

\section{\label{sec:summ}Conclusion}

%
\begin{figure}
\begin{centering}
\includegraphics[width=0.45\textwidth]{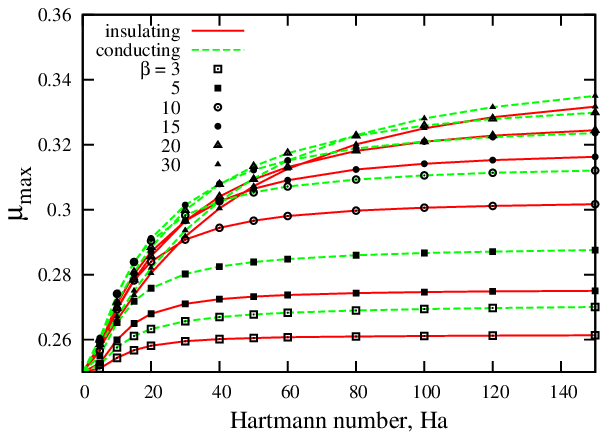} 
\par\end{centering}

\caption{\label{cap:muc-Ha} Maximal value of $\mu$ versus the Hartmann number
for absolute instability at various $\beta$.}

\end{figure}

In this study we have analyzed numerically the MRI of Taylor-Couette
flow with a helical external magnetic field. The problem was considered
in the inductionless approximation defined by a zero magnetic Prandtl
number ($\Pm=0).$ First, we carried out a conventional linear stability
analysis for perturbations in the form of Fourier modes specified
by real wave numbers. The helical magnetic field was found to extend
the original instability to a relatively narrow range beyond its purely
hydrodynamic limit defined by the Rayleigh line. The range of destabilization
was found to be considerably larger for perfectly conducting cylinders
than that for insulating ones. For insulating cylinders, the instability
beyond the Rayleigh line is effective only in a limited range of wave
and Reynolds numbers. Unstable Reynolds numbers are bounded by an
upper critical value which tends to infinity right beyond the Rayleigh
line. For perfectly conducting cylinders and moderate helicities of
the magnetic field, the range of unstable wave numbers is bounded
only from the short-wave end. Although there is an upper marginal
Reynolds number for each unstable wave number, no bounded upper critical
Reynolds number exists in this case because the range of unstable
wave numbers extends to zero, \textit{i.e.}, infinitely long waves.
Nevertheless, at sufficiently large helicities, the range of unstable
wave numbers becomes bounded also from below, and an upper critical
Reynolds number appears in the same way as for insulating cylinders.

It is important to note that these instabilities predicted by the
conventional stability analysis in the form of single traveling waves
correspond to the so-called convective instability threshold at which
the system becomes able to amplify certain externally imposed perturbations
that, however, are not self-sustained and thus may be experimentally
unobservable without a proper external excitation. The problem is
that convectively unstable perturbations grow asymptotically in time
only in the frame of reference traveling with their group velocity,
whereas they decay in any other frame of reference including the laboratory
one. For an instability to be self-sustained and thus observable it
has to grow in the laboratory frame of reference. In an extended system,
this condition is satisfied by the so-called absolute instability
which ensures a zero group velocity of a growing perturbation. This
additional condition is satisfied by regarding the wave number as
a complex quantity with a nonzero imaginary part which describes an
exponential axial modulation of the wave amplitude. Using this concept,
we found that there is not only a convective but also an absolute
HMRI implying that this instability can be experimentally observable
in a system of sufficiently large but finite axial extension. In the
hydrodynamically unstable range before the Rayleigh line, the threshold
of absolute instability is slightly higher than the convective one.
Nevertheless, the critical wavelength for absolute instability is
significantly shorter than that for the convective one that may allow
us to distinguish between both. The absolute instability threshold
rises significantly above the convective one beyond the Rayleigh line.
As a result, the extension of the absolute instability beyond the
Rayleigh line is considerably shorter than that of the convective
instability without a marked difference between insulating and perfectly
conducting cylinders in contrast to the convective HMRI.

The extension of HMRI beyond the Rayleigh line is of particular interest
from the astrophysical point of view regarding a Keplerian velocity
profile \cite{Liu-etal2006,Ruediger-Hollerbach2007}. For a Couette-Taylor
flow with a radius ratio $\lambda=2$ considered here, the Keplerian
velocity profile approximately corresponds to a ratio of rotation
rates of $\mu=\lambda^{-3/2}\approx0.35.$ As seen in Figs. \ref{cap:muc-bt}(b)
and \ref{cap:muc-Ha} for the absolute instability, no such value
of $\mu$ is reached up to $\beta=30$ and $\Ha=150$. Whether or
not it can be reached at higher $\beta$ and $\Ha$ is still an open
question requiring a more detailed study using either higher numerical
resolution or asymptotic analysis. On the other hand, HMRI in a system
of large axial extension with a radius ratio of $\lambda=2$ might
be of limited astrophysical relevance for accretion disks anyway.

Conversely to Liu \emph{et al.} \cite{Liu-etal2006,Liu-etal2007}
we find that the HMRI can be a self-sustained instability rather than
just a transient growth. This contradiction may be due to a couple
of additional simplifications underlying the analysis of Liu \emph{et
al.} First, the viscosity is neglected. Second, the electromagnetic
force is treated as a small perturbation which is a sensible approach
within the inviscid approximation where an infinitesimal magnetic
field can cause a correspondingly slow growth of an unstable Fourier
mode. However, such a perturbative approach may be inadequate for
the absolute instability which requires a finite temporal growth rate
of the corresponding convectively unstable Fourier mode. In addition,
note that our results do not support the recent findings of Liu \cite{Liu-2009}.
According to his estimates for the cases considered here, the absolute
HMRI occurs above $\RE\sim10^{5}$ only. We predict the absolute HMRI
to occur at the lower value of $\RE\sim10^{3},$ and to disappear
again above $\RE\sim10^{5}$ due to the mechanism discussed in Sec.
\ref{sub:conv-insl}. This disagreement may be due to the absolute
instability analysis which is carried out by Liu using real wave numbers
only. According to the conventional absolute instability analysis
\cite{Lifshitz-Pitaevskii-81,Schmid-Henningson}, such an approach
yields the long-time asymptotics only in the frame of reference traveling
with the group velocity of the fastest growing perturbation. Moreover,
owing to its linearity, the analysis is limited to sufficiently small
perturbation amplitudes only. Namely, it is limited up to the point
where the first exponentially growing perturbation appears, which
is the convective instability threshold.

%
\begin{figure*}
\begin{centering}
\includegraphics[width=0.45\textwidth]{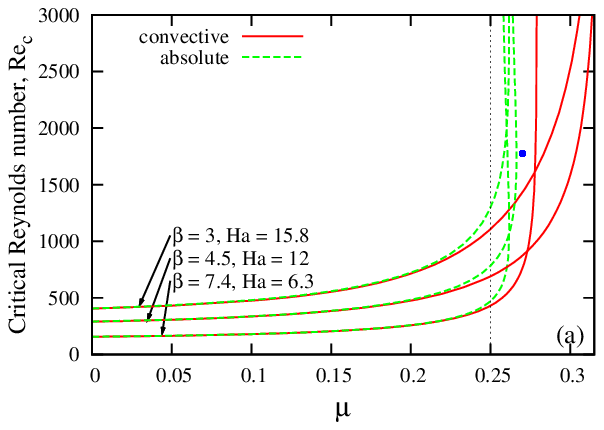}\includegraphics[width=0.45\textwidth]{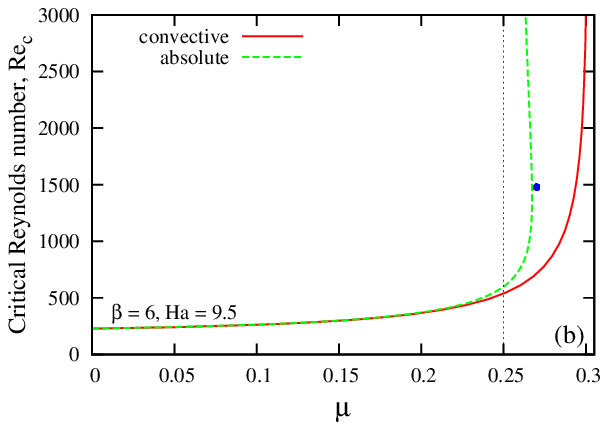} 
\par\end{centering}

\caption{\label{fig:rec-mu-exp} Comparison of the critical Reynolds number
$\RE_{c}$ for convective and absolute instability thresholds in the
case of perfectly conducting cylinders with the experimental values
(dots) for which traveling waves have been observed at (a) $\mu=0.27$,
$\RE=1775,$ $\beta\approx7.4-3,$ and $\Ha\approx6.3-15.8$ and (b)
$\RE=1479$, $\beta=6$, and $\Ha=9.5.$ }

\end{figure*}

Finally, let us compare the convective and absolute instability thresholds
calculated for perfectly conducting cylinders with the experimental
data of Stefani \emph{et al.} \cite{Stefani-NJP} who reports on
the observation of HMRI-like traveling waves at $\mu=0.27,$ $\RE=$1775,
and a fixed rod current $6\, kA$ for the coil currents $40-100\, A$
that corresponds to $\beta\approx7.4$--3 and $\Ha\approx6.3$-15.8.
Another observation was done at the same $\mu$ but different other
parameters: $\RE=1479,$ $\beta=6,$ and $\Ha=9.5.$ As seen in Fig.
\ref{fig:rec-mu-exp}, in both cases the experimental points lie well
inside the range of $\mu$ for convective instability but outside
that for absolute instability. This discrepancy with the experimental
observations may be due to the deviation of the real base flow from
the idealized one used in this study. In particular, the Ekman pumping
driven by the end walls in the experiment, which is not taken into
account in the present analysis, may affect the hydrodynamic stability
limit of the base flow, \textit{i.e.}, its actual Rayleigh line, which
however serves as the reference point for the observation of MRI.
A more detailed comparison with the experimental observations lies
outside the scope of the present paper.

In conclusion, the main result of the present paper is the finding
of absolute HMRI in addition to the convective one which can be self-sustained
and, thus, experimentally observable without external excitation in
a system of sufficiently large axial extension. A characteristic feature
of HMRI is the upper critical threshold existing besides the lower
one that distinguishes it from a magnetically-modified Taylor vortex
flow.

\begin{acknowledgments}
This research was supported by Deutsche Forschungsgemeinschaft in
frame of the Collaborative Research Centre SFB 609. We would like
to thank Frank Stefani and Thomas Gundrum for stimulating discussions. 
\end{acknowledgments}

\end{document}